# Vertical graphene base transistor

W. Mehr[a*], J. Ch. Scheytt[a], J. Dabrowski[a], G. Lippert[a], Y.-H. Xie[b], M. C. Lemme[c], M. Ostling[c], G. Lupina[a]

[a]IHP, Im Technologiepark 25, 15236 Frankfurt (Oder), Germany

[b]Department of Materials Science and Engineering, University of California at Los Angeles, Los Angeles, CA 90095-1595 USA

[c]KTH Royal Institute of Technology, Isafjordsgatan 22, 16440 Kista, Sweden.

*E-mail: mehr@ihp-microelectronics.com

## Abstract

We present a novel, graphene-based device concept for high-frequency operation: a hot electron graphene base transistor (GBT). Simulations show that GBTs have high current on/off ratios and high current gain. Simulations and small-signal models indicate that it potentially allows THz operation. Based on energy band considerations we propose a specific materials solution that is compatible with SiGe process lines.

## I. Introduction

Carbon-based materials may enhance the performance of digital and radio frequency (RF) electronics [1]. Extensive research has been devoted to exploit the exceptional properties of graphene in field-effect transistors with a graphene channel (GFETs) [2, 3]. This has resulted in RF GFETs with transition frequencies ($f_T$) of several hundred GHz [4, 5] ambipolar RF mixers [6], and frequency multipliers [7]. However, GFETs are not suitable for logic applications due to the absence of a band gap, and the lack of pronounced drain current saturation limits their potential for conventional RF amplifying circuits [8, 9].

We propose an alternative application of graphene as an extremely thin, highly conductive electrode for a hot electron transistor with a graphene base. This graphene base transistor (GBT) combines the concept of hot electron transistors (HET) [10-12] with the unique properties of graphene to result in a high frequency device that offers low off currents ($I_{off}$), drain current saturation and power amplification.

## II. Graphene base transistor concept

Figures 1a and b illustrate the difference between the GFET and the GBT. Charge carriers traverse the graphene in the GFET laterally. The GBT is based on a vertical arrangement of emitter (E), base (B), and collector (C), just like a hot electron transistor or a vacuum triode. In the off-state, the carriers face a barrier (cf. the simplified band diagram in Fig. 1c). Note that although graphene has no band gap for lateral transport, it poses a barrier for transport in the normal direction (band gap at $\Gamma$ [13]). In the on-state, cariers tunnel through the emitter-base insulator (EBI) and the base control electrode (graphene) into the conducting band of the base-collector insulator (BCI; Fig. 1d).

There are substantial advantages when using graphene as the base material: The monatomic thickness favors ballistic transport across the base and a homogenous electric field at the base interface. Assuming that only electrons scattered within the base contribute to the base current $I_B$, this should reduce $I_B$ by two orders of magnitude compared to a similar HETs with a metal base. In contrast to metals, the base resistivity is not limited by pinholes; values around 100 $\Omega$/sq are achievable [14]. Graphene is chemically inert, reducing issues with process-induced interface reactions. Although inhomogeneity of graphene doping may lead to inhomogeneous $I_C$ and to local heating, this may be uncritical thanks to high thermal conductivity of graphene.

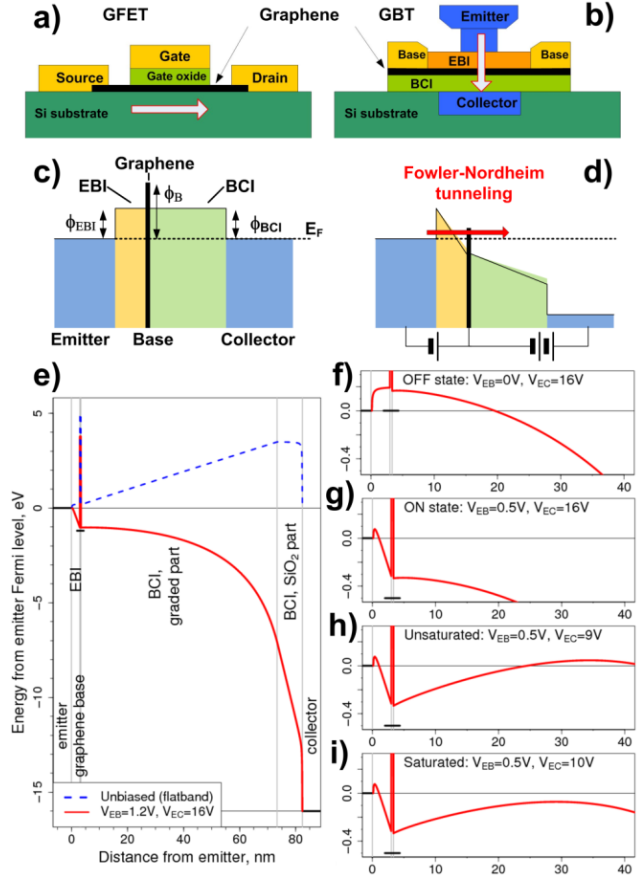

Figure 1. (**a-d**) Schematic cross-sections of (**a**) GFET and (**b**) GBT and schematic band diagrams of (**c**) an unbiased and (**d**) biased GBT. (**e-i**) Calculated band diagrams of a high power GBT with graded BCI. $V_{EB}$ is given with respect to bias compensating the work function difference between graphene and emitter (flatband). (**e**) Potential distribution at flatband (blue dashed) and at $V_{EB}$=1.2V, $V_{EC}$=16V (red solid). (**f-i**) Potential distribution close to graphene. At $V_{EC}$=16V: (**f**) OFF state, (**g**) ON state. At $V_{EB}$=0.5V: (**h**) unsaturated regime with potential barrier in the BCI and (**i**) saturated regime.

### III. Basic design aspects

GBT needs to be carefully engineered for optimal operation. The EBI must be thin to yield high output currents. The EBI barrier $\Phi_{EBI}$ is controlled by the E-B voltage ($V_{EB}$) applied to the graphene. In the ON state, electrons must cross the BCI easily. However, for good power performance, the BCI should withstand $V_{BC} \approx$ 10 V, which implies a high tunneling barrier. The structure shown in Fig. 1e addresses these issues. SiO$_2$ is used on the collector side and a graded silicate on the base side. In the graded part, the dielectric constant and the BCI barrier ($\Phi_{BCI}$) vary with the distance from the base. This can be achieved with a gradually decreasing metal content across the dielectric [15]. The barrier on the collector side is controlled by the B-C voltage $V_{BC}$. When $V_{BC}$ is high enough, most of the electrons encounter no barrier (Fig. 1e). SiO$_2$ thickness allows for high output voltages, i.e. for good power performance. This reveals several advantages of the GBT compared to GFETs: (a) GBTs allow for high $I_{on}/I_{off}$ current ratios (Fig. 3b-c); (b) GBTs show current saturation in the output characteristics, because for high $V_{EC}$, nearly all electrons travel above the BCI barrier. Thus, $I_{on}$ is limited by the EBI barrier and independent of $V_{EC}$ (Fig. 1i); (c) Tunneling is a fast process. Even at 2.5 THz, the current response of a tunneling diode resembles the dc curve. With the transport distance below 100 nm, delays due to diffusion should stay below a picosecond [16].

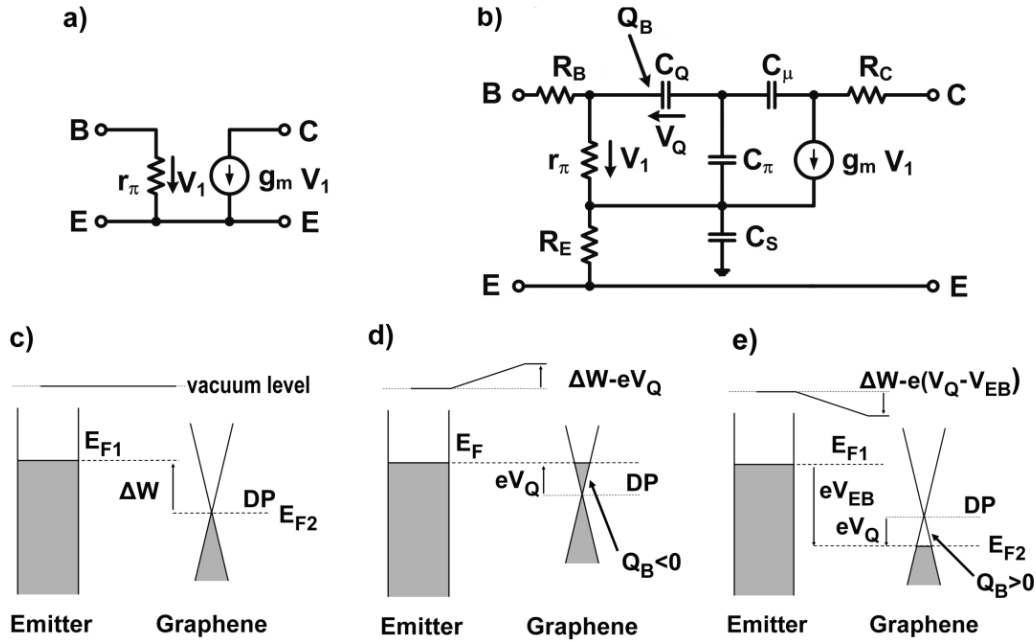

Figure 2. (**a-b**) Small-signal models at (**a**) low and (**b**) high frequency. (**c-d**) Quantum capacitance: (**c**) before and (**d**) after equilibration of Fermi levels, and (**e**) under bias $V_{EB}$. $\Delta W \approx 0.6$ eV, $\Delta W - eV_Q \approx 0.3$ eV

## IV. Small-signal modelling and comparison with HBT

A THz transistor may work as a high-frequency (HF) linear small-signal (SS) amplifier. SS models with and without parasitics are given in Fig. 2a-b. The transconductance $g_m$ becomes:

$$g_m = \frac{\partial i_C}{\partial v_1} = \frac{\beta_0}{\beta_0 + 1}\frac{\partial i_E}{\partial v_1} \approx \frac{\partial i_E}{\partial v_1}, \qquad (1)$$

where $v_1$ is the SS voltage (i.e., $V_{BE,ac} = V_{BE,dc} + v_{1,ac}$ and the amplitude of the ac signal $v_1$ is small), $i_c$ and $i_b$ are the SS collector and base currents, $\beta_0 = i_c / i_b \gg 1$ is the SS current gain.

The HF-SS model (Fig. 2b) assumes metallic emitter and collector, and graphene base. $R_B$ denotes the resistance of the base contact and of the graphene layer, $R_C$ and $R_E$ represent the collector and emitter resistance, $r_\pi$ and $r_\mu$ are the differential resistances of EBI and BCI, $C_\pi$ and $C_\mu$ are their plate capacitances. $C_Q = |\partial Q_B/V_Q| = \kappa |V_Q|$ is the quantum capacitance of graphene, $\kappa = 25$ µF/cm$^2$/V [17]. $Q_B$ is the charge accumulated in graphene and $eV_Q$ has the physical meaning of the Fermi energy in graphene, measured with respect to the Dirac point (DP in Fig. 2c-e). Neglecting the substrate capacitance $C_s$ and delays due to diffusion of carriers, the frequency response is:

$$\tau = \frac{dQ_B}{di_C} = \frac{C_{TOT}}{g_m}; \quad f_T = \frac{1}{2\pi\tau} = \frac{1}{2\pi}\frac{g_m}{C_{TOT}}. \qquad (2)$$

$$C_{TOT} = \frac{C_Q(C_\pi + C_\mu)}{C_Q + C_\pi + C_\mu}, \qquad (3)$$

$$V_Q = \frac{C_\pi + C_\mu}{s\kappa}\left(\sqrt{1 + 2\kappa\frac{|C_\pi U_{EB} + C_\mu U_{CB}|}{(C_\pi + C_\mu)^2}} - 1\right), \qquad (4)$$

with $s = \text{sign}(C_\pi U_{EB} + C_\mu U_{CB})$, $U_{EB} = V_{EB} - \Delta W$, $U_{CB} = V_{CB} + \Delta W$, and $V_{CB} = V_{EB} - V_{EC}$. The accumulated charge is $Q_B = \frac{1}{2} s \kappa V_Q^2$. For metallic base, one obtains $C_{TOT} = C_\pi + C_\mu$ [18]; $V_Q$ is then 0.

## V. Quantum mechanical simulations

Graphene is semi-metallic, with the Dirac point in the corner of the two-dimensional first Brillouin zone (high lateral momentum). In the GBT, electrons tunnel through the EBI and most of them are likely to enter graphene with small lateral momentum. For such electrons there is an energy gap in graphene (at $\Gamma$). To verify if this makes graphene a tunneling barrier, we simulated the tunneling across graphene placed between unbiased cobalt electrodes in vacuum. The selfconsistent band structure obtained from ab initio atomistic calculations was used. Insertion of graphene between the electrodes separated by 1.9 nm results in a tunneling spectrum roughly proportional to that obtained for 1.7 nm separation and no graphene. Thus, graphene slightly reduces the vacuum barrier strength. This is largely due to work function difference between cobalt and graphene: graphene becomes positively charged, so that the distance to vacuum energy decreases as the electron approaches the graphene sheet. A fully transparent or scattering-only sheet would reduce the vacuum thickness by about 0.35 nm (i.e., by the thickness of graphene) even without any work function difference. This does not happen; hence graphene is a barrier, not a transparent layer.

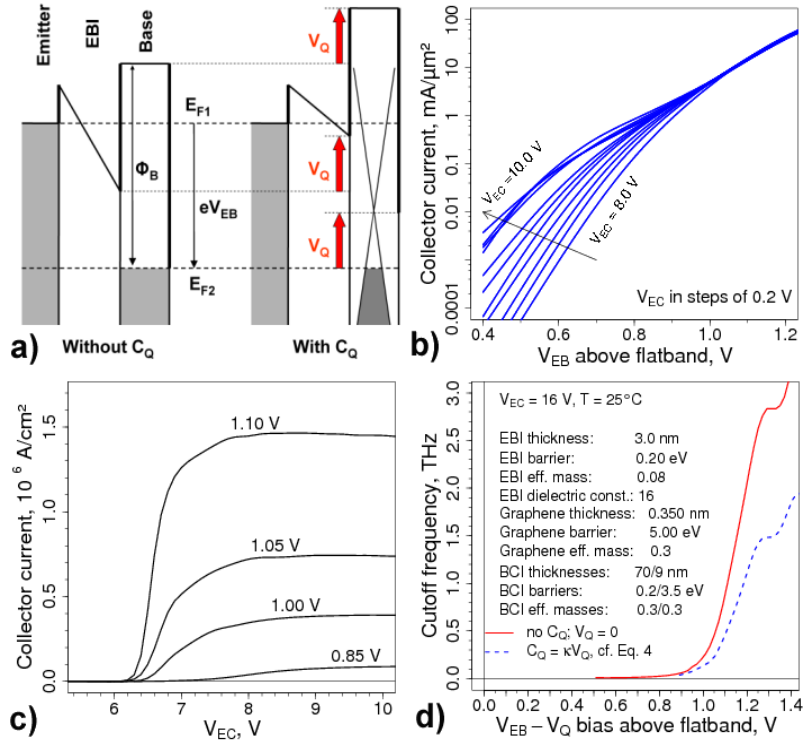

Figure 3. (**a**). The effect of $C_Q$ on the EBI electric field and on the effective $\Phi_B$. (**b**) Transfer characteristics for common emitter operation. (**c**) Output characteristics for various $V_{EB}$. (**d**) Transition frequency $f_T$ obtained without $C_Q$ effects (solid) and with $C_Q$ (broken, cf. Eq. 2) is plotted against $V = V_{EB} - V_Q$, as this defines the EBI electric field. $V_Q \approx 0.3$ V for $V \approx 1.3$ V. Above 1.2 eV, quantum oscillations in $f_T$ begin.

We performed quantum-mechanical simulation of the GBT. The tunneling parameters cannot be derived reliably from our atomistic data. For that reason the simulation should be viewed as a zero-order estimate of a GBT in action. The Schrödinger equation with open boundary conditions was solved numerically for one-band effective potential rounded up by image force at interfaces with emitter and collector. No self-consistent term was added, as the distribution of the potential in vicinity of graphene is not known exactly. No scattering effects were included; the temperature corresponds to the Fermi distribution of electron energies. We approximate the tunneling barrier as a rectangle with $d = 0.35$ nm and $\Phi_B = 5$ eV (the conduction band edge at $\Gamma$ is between 3.7 eV and 7 eV [19, 20]). The effective mass was set to 0.3, a conservative value typical for, e.g., $SiO_2$. We performed the calculations with and without quantum ca capacitance effects. Quantum capacitance lowers $C_{TOT}$, increasing $f_T$. But in a realistic device $-C_\pi U_{EB}$ will exceed $C_\mu U_{CB}$, hence $Q_B > 0$. This reduces the electric field in EBI and increases the effective $\Phi_B$ (Fig. 3a), decreasing $g_m$ and thus also $f_T$. With increasing $\Delta W$ the $f_T$ degradation becomes less pronounced.

Figure 3b-c shows the simulated transfer and output characteristics for operation as a power amplifier. These curves underscore the potential of the GBT, with the $I_C$ switching over several orders of magnitude (Fig. 3b) and $I_C$ saturation (Fig. 3c). We estimate that for THz operation (Fig. 3d) at $V_{EB} \approx 1$V, the EBI should be not thicker than 3-5 nm and its energy barrier $\Phi_{EBI}$ at no bias should be 0.4 eV or less. In our estimations with $V_{EC} = 16$ V, the electric field in $SiO_2$ is close to the critical field and below the critical field in the rest of the BCI. Unpinned $Er_2Ge_3$/Ge is assumed for the emitter/EBI. This should be viable as the interface between Ge and a germanide can be unpinned by, e.g., P [21]. The work function of $Er_2Ge_3$, 4.05 eV, matches the electron affinity of Ge, 4.0 eV. Assuming that the $Er_2Ge_3$/Ge interface can be unpinned as efficiently as for PrGe/Ge, we take $\Phi_{EBI} = 0.2$ eV at no bias. For the graded part of the BCI we use $Ti_xSi_{1-x}O_2$. The barrier at graphene/$TiO_2$ is assumed the same as at Ge/graphene [22]. Figure 3d compares $f_T$ obtained without $C_Q$ influence ($C_Q \to \infty, V_Q \to 0$) and with $C_Q$ (using $\Delta W = 0.6$ V).

## VI. Conclusions

A new device, a graphene base transistor, GBT, has been proposed and analyzed. The key feature is the use of graphene as the base electrode in a hot electron transistor configuration. Distinct advantages are that graphene is pinhole-free and does not interact chemically with adjacent materials. Graphene is also a highly conductive, one-atom thick film, which does not scatter the electrons injected from the emitter to the base. Simulated GBT transfer characteristics show switching over several orders of magnitude and output characteristics show clear saturation. We proposed and evaluated a specific materials solution for GBT indicating feasibility of THz operation.

**Acknowledgement**

We thank F. Driussi, P. Palestri, and L. Selmi for discussions. Atomistic calculations have been done at the Jülich Supercomputing Centre, Germany, NIC project hfo06. M.C. Lemme and M. Ostling acknowledge support through an Advanced Investigator Grant (OSIRIS, No. 228229) from the European Research Council.